\documentclass{aa}
\usepackage{graphicx}

\newcommand{\alms}{a_{lm}'s}
\newcommand{\LA}{likelihood analysis}

\newcommand{\vd}{\overrightarrow{d}}

\newcommand{\vtc}{\overrightarrow{\tilde{c}}}

\newcommand{\va}{\overrightarrow{a}}

\newcommand{\vf}{\overrightarrow{f}}

\newcommand{\vTheta}{\overrightarrow{\Theta}}

\newcommand{\mY}{\vec{Y}}

\newcommand{\mPsi}{\vec{\Psi}}
\newcommand{\vTeta}{\overrightarrow{\Theta}}
\newcommand{\mC}{\vec{C}}
\newcommand{\mT}{\vec{T}}
\newcommand{\mI}{\vec{I}}
\newcommand{\mtT}{\vec{\tilde{T}}}
\newcommand{\mtN}{\vec{\tilde{N}}}

\newcommand{\mN}{\vec{N}}
\newcommand{\mA}{\vec{A}}
\newcommand{\mW}{\vec{W}}
\newcommand{\mM}{\vec{M}}

\newcommand{\mU}{\vec{U}}
\newcommand{\mGam}{\vec{\Gamma}}
\newcommand{\hn}{\hat{n}}

\newcommand{\Npix}{N_{\rm pix}}
\newcommand{\Ompix}{\Omega_{\rm pix}}
\newcommand{\fbP}{\delta T_{\rm fb}}

\newcommand{\dTnorm}{\delta T_{\rm N}}
\newcommand{\ket}{> }
\newcommand{\bra}{< }
\newcommand{\Omo}{\Omega}
\newcommand{\Omb}{\Omega_{\rm b}}
\newcommand{\Omc}{\Omega_{\rm cdm}}
\newcommand{\Omk}{\Omega_{\rm k}}
\newcommand{\OmL}{\Omega_\Lambda}
\newcommand{\Ho}{H_{\rm o}}
\newcommand{\GOF}{Goodness of Fit}

\begin{document}

%
   \title{Concerning Parameter Estimation Using the Cosmic 
        Microwave Background}
%

        \titlerunning{About ...} 

   \author{ M.~Douspis$^1$, J.G.~Bartlett$^1$, A.~Blanchard$^{1,2}$ 
        \& M.~Le~Dour$^1$}


   \institute{$^1$ Observatoire Midi-Pyr\'en\'ees,
              14, ave. E. Belin,
              31400 Toulouse, FRANCE \\
              Unit\'e associ\'ee au CNRS 
             ({\tt http://www.omp.obs-mip.fr/omp})\\    
              $^2$ Observatoire de Strasbourg,
                   Universit\'e Louis Pasteur,
                   11, rue de l'Universit\'e,
                   67000 Strasbourg,
                  FRANCE\\
                   Unit\'e associ\'ee au CNRS 
                   ({\tt http://astro.u-strasbg.fr/Obs.html})
             }

   \date{November, 2000}

   \maketitle

   \begin{abstract}
     Most parameter constraints obtained from 
cosmic microwave background (CMB) anisotropy data 
are based on power estimates and 
rely on approximate likelihood functions; 
computational difficulties generally preclude an exact
analysis based on pixel values.  With the
specific goal of testing this kind of
approach, we have performed a complete (un-approximated)
likelihood analysis combining the  
COBE, Saskatoon and MAX data sets.  We examine in
detail the ability of certain approximate techniques 
based on band--power estimates to recover the 
full likelihood constraints.  The traditional $\chi^2$--method 
does not always find the same best--fit model as the 
likelihood analysis (a bias), due mainly  
to the false assumption of Gaussian likelihoods that
makes the method overly sensitive to data outliers.  
Although an improvement, other approaches employing 
non--Gaussian flat--band likelihoods do not always 
faithfully reproduce the complete likelihood constraints either; 
not even when using the exact flat--band likelihood curves.  
We trace this to the neglect of spectral 
information by simple flat band--power estimates.
A straightforward extension incorporating a local 
effective slope (of the power spectrum, $C_l$) provides a faithful 
representation of the likelihood surfaces without 
significantly increasing computing cost.
Finally, we also demonstrate that the best--fit model to this
particular data set is a {\em good fit}, or that the observations 
are consistent with Gaussian sky fluctuations, according
to our statistic.
      \keywords{cosmic microwave background -- Cosmology: observations --
        Cosmology: theory}
   \end{abstract}


\section{Introduction}

        The extraction of information from cosmic microwave
background (CMB) anisotropies is a classic problem of 
model testing and parameter estimation, the goals
being to constrain the parameters of 
an assumed model and to decide if
the {\em best--fit model} (parameter values) 
is indeed a good description of the data. 
Maximum likelihood is often used as the method
of parameter estimation.  Within the context
of the class of models to be examined, the probability 
distribution of the data 
is maximized as a function of the model parameters,
given the actual, observed data set.  This is the same as a 
Baysian analysis with uniform priors.  Once found,
the best model must then be judged on its ability to
account for the data, which requires
the construction of a {\em statistic} 
quantifying the {\em goodness--of--fit} (GoF).
Finally, if the model is retained as a good fit,
one defines {\em confidence} intervals on the
parameter estimation.  The exact meaning of these
confidence intervals depends heavily on the method
used to construct them, but the desire is 
always the same -- one wishes to quantify
the `ability' of other parameters to explain
the data (or not) as well as the best fit values.

     Data on the CMB consists of sky brightness
measurements, usually given in terms of equivalent
temperature.  An experiment may produce a true
map, for example, the COBE maps, or a 
set of temperature differences, such as published
by the Saskatoon experiment.  The
likelihood function is to be constructed using
these pixel values\footnote{The term
pixel will be understood to also include temperature
differences.}.  Standard Inflationary scenarios predict
{\em Gaussian} sky fluctuations, which implies that
the pixels should be modeled as random variables 
following a multivariate normal distribution, with 
covariance matrix given as a function of the model 
parameters (in addition to a noise term).  
It is important to note that, since the parameters enter 
through the covariance
matrix, and not as some linear combination of 
pixel values, {\em the likelihood function is 
{\bf not} Gaussian}.

     Although it would seem straightforward to 
estimate model parameters directly with
the likelihood function, in practice the procedure is
considerably complicated by the complexity
of the model calculations and by the size of the data sets
(Bond et al. 1998, 2000; Borrill 1999ab; Kogut 1999).
Maps consisting of several tens of thousands of pixels
(the present situation) are extremely cumbersome
to manipulate\footnote{Note that even the recent
analyzes of the BOOMERanG and MAXIMA--1 data sets
relied on approximate methods (Balbi et al. 2000;
Jaffe et al. 2000; Lange et al. 2000).}, and 
the million--pixel maps expected from MAP and Planck 
cannot be analyzed by this method in any practical way.  
An alternative is to first estimate the angular power
spectrum from the pixel data and then work with this
reduced set of numbers.  For Gaussian fluctuations,
there is in principle no loss of information.  
Because of the large reduction of the data ensemble to 
be manipulated, the tactic has been referred to 
as ``radical compression'' (Bond et al. 2000).  
The power spectrum
has in fact become the standard way of reporting CMB results;
it is the best visual way to understand the data,
and in any case it is what is actually calculated 
in the models.  

     The critical issues are then how to best
go from the pixel representation to the power spectrum,
and how to correctly use the power spectrum for
parameter estimation and model evaluation.  
Consider the former issue.
Because any given experiment has only limited 
spatial frequency resolution, due to incomplete sky coverage, 
one may only obtain the signal power over
a finite range (or band) of multipoles. 
Extraction of this power is itself
a question of parameter estimation, where
the parameter is simply the {\em band--power}.
Band--powers are thus themselves commonly found by
using the likelihood.  For first
generation experiments\footnote{We use the
term first generation to refer to 
COBE and experiments prior to BOOMERanG 
and MAXIMA--1; apart from COBE, most of 
these were not optimized for map construction, but
rather power extraction over
a limited range of scales.}, band--powers
and their uncertainties can be found
by completely mapping--out the band--power
likelihood function.  The large data sets 
from, for example, BOOMERanG (de Bernardis et al.
2000) and MAXIMA (Hanany et al. 2000), 
preclude this possibility and require 
approximate likelihood methods: for instance,
one first determines the best--fit band--powers 
by just finding the maximum of the likelihood function;
the shape of the likelihood around the maximum
is then modeled with a well--motivated but
approximate expression (Bond et al. 2000).

     Consider now the second issue.
Almost without exception, present efforts to 
constrain cosmological parameters use
these band--power estimates as their
starting point.  In addition, many  employ
a simple $\chi^2$ minimization over
the power points with the supplied `error bars'. 
There are two relevant remarks to make:
firstly, the $\chi^2$ method is not appropriate for the
task because band--power estimates
do not represent Gaussian distributed data.
Secondly, the `error bars' are most often
defined in a Baysian fashion by treating the 
band--power likelihood function as a probability 
distribution; these do not necessarily
represent the error distribution of the 
power estimator (defined as the expression
maximizing the likelihood) that is
required by the $\chi^2$ method, and
which a frequentist would argue must be found 
by simulation.

     These comments motivate a more in depth consideration
of the general problem of parameter estimation
using band--powers.  Some authors have
attempted to construct simple approximations
to the band--power likelihood function that
only require limited information, such as the
best power estimate and associated Baysian
confidence limits (Bartlett et al. 2000, Bond et al. 2000,
Wandelt et al. 2000).  A likelihood analysis
over physical parameters (e.g., $\Omo$)
then follows by inserting the model
dependence of the band--power into the
approximate function.  
This kind of approach has a two--fold use: firstly,
it permits one to analyze the ensemble of first generation
data (e.g., Le Dour et al. 2000), and secondly, 
it permits one to analyze
larger data sets by just finding the best fit
band--powers (e.g., Jaffe et al. 2000).
Beyond the question
of the accuracy of the approximation,
one may worry that perhaps the whole approach
is insufficient -- that, even with the
exact band--power likelihood function,
the parameter constraints would not
correspond to the more complete likelihood
analysis using the entire pixel set.  
The fact that in principle the power spectrum
contains the same information as the sky
temperatures (at least for Gaussian fluctuations)
does not guarantee that band--powers 
do as well; this depends on their definition,
which usually adopts a 
certain spectral form.  The common {\em flat 
band--powers} are defined by the assumption
of a flat spectrum over the band.  This may lead 
to a concern that important information on
the slope of the spectrum is lost, a perhaps
extremely relevant issue around the Doppler 
peaks.

     The goal of this {\em paper} is to examine
some of these questions by performing a complete
likelihood analysis (in pixel space) of a
subset of present CMB data; in particular,
we study the information content of flat band--power 
compression and the validity of likelihood
approximations.  The data subset
consists of the COBE, Saskatoon (CAP) and MAX 
experimental results.  Taken together, they cover 
a wide range of angular scales, including
the region of the first (so--called) Doppler peak predicted by
Inflation models, and thereby allow non--trivial 
parameter constraints to be established.
Our complete analysis permits
us to evaluate the performance of power--plane
methods, such as  
$\chi^2$ minimization and the recently
proposed band--power likelihood approximations
(Bartlett et al. 2000, Bond et al. 2000, Wandelt et al. 2000).
We find that the $\chi^2$ approach is overly
sensitive to outliers because of its incorrect
assumption of normal distributions; we shall
see an example where this leads to a bias
in the deduced best--fit model parameters.
The approximate methods perform better, but
even here we find some significant differences
with the full analysis.  These differences
remain even if we simply interpolate the
band--power likelihood functions, something that
leads us to discover that the
data set is not fully described by just
its set of band--powers.  The experiments
are in fact individually somewhat sensitive to the
shape of the spectrum, and the  
power estimates therefore depend on more than just
the total in--band power.  In most cases
the experimental power is adequately 
modeled as a function of the normalization
and local slope of the spectrum.  

     Finally, we address the commonly neglected
issue of the \GOF\ (GoF)of the best model.  We
demonstrate that the best--fit model to this 
data set is indeed a {\em good} fit.  This may
also be interpreted as saying that our \GOF\ 
statistic is consistent with
Gaussian temperature fluctuations.


\section{Likelihood Method}

        The common approach to CMB data analysis is through
the likelihood function, ${\cal L}$.  Given a set of pixels,
this function relates the prediction of a particular model
to observations, taking into account, for example,
the effects of beam smearing and the observing strategy.
We will analyze in this section three experiments 
within the likelihood framework: MAX (Clapp et al. 1994, Tanaka et al. 1996, Ganga et al. 1998), Saskatoon (Netterfield et al. 1997) 
and the COBE 4--year maps (Bennett et al 1996).  
Each presents a different 
type of observing strategy.  These three experiments
are sensitive to different angular scales - schematically
COBE provides information on the amplitude of the power 
spectrum, while MAX and Saskatoon tell us about the 
position and height of the first acoustic peak.  
They hence prove quite
complimentary in a likelihood analysis on
cosmological parameters.  The results of such an analysis
for open Inflationary models are given at the end of 
this section.  They will subsequently be used as a benchmark
against which we will be able to test various approximate
methods.

\subsection{Generalities}

Temperature fluctuations of the CMB are described by a random 
field in two dimensions:  $\Delta(\hat{n})\equiv (\delta T/T) (\hat{n})$, 
where $T$ refers to the temperature of the background and 
$\hat{n}$ is a unit vector on the sphere. It is usual to 
expand this field using spherical harmonics: 
\begin{equation}\label{eq:alm}
\Delta(\hat{n}) = \sum_{lm} a_{lm} Y_{lm}(\hat{n})
\end{equation}
The $a_{lm}$'s are randomly selected from the probability distribution 
characterizing the process generating the perturbations. In the 
Inflation framework, which we consider in this paper, the $a_{lm}$'s  are 
{\em Gaussian random variables} with zero mean and covariance
\begin{equation}\label{eq:defCl}
<a_{lm}a^*_{l'm'}>_{ens} = C_l \delta_{ll'}\delta_{mm'}
\end{equation}                
The $C_l$'s then represent the {\em power spectrum}.   
We may express the correlation between two points separated on
the sky by an  angle $\theta$ as
\begin{equation}
C(\theta) \equiv <\Delta(\hat{n}_1)\Delta(\hat{n}_2)>_{ens} =
        \frac{1}{4\pi}\sum_{l} (2l+1) C_l P_l(\mu)
\end{equation}  
where $P_l$ is the Legendre polynomial of order $l$
and $\mu = \cos\theta = \hat{n}_1\cdot\hat{n}_2$. The statistical isotropy of
the perturbations demands that the correlation function depend only on
separation, $\theta$, which is in fact what permits such an expansion.

When observed, the temperature fluctuations are convolved by the 
experimental beam, $B$, positioned on the sky at $\hat{n}_p$:
\begin{equation}
\Delta_b(\hat{n}_p) = \int\; d\Omega \Delta(\hat{n}) B(\hat{n}_p,\hat{n})
\end{equation} 
If the beam can be described by harmonic coefficients, $B_l$, defined by
\begin{equation}\label{eq:Bl}
B(\theta') = \frac{1}{4\pi}\sum_l (2l+1) B_l P_l(\mu')
\end{equation} 
where $\hat{n}_p\cdot\hat{n} = \cos\theta' = \mu'$, 
then the observed (or beam-smeared) correlation function may simply be 
calculated as a convolution on the sphere:
\begin{eqnarray}\label{eq:Cbeam}
C_{b}(\theta) \equiv <\Delta_b(\hat{n}_1)\Delta_b(\hat{n}_2)>_{ens}
        & = &\frac{1}{4\pi} \sum_l (2l+1) C_l \\
\nonumber
& & \hspace*{1cm} \times |B_l|^2 P_l(\mu)
\end{eqnarray}   
Note that expansion ($\ref{eq:Bl}$) pre--supposes axial symmetry for the
beam.

        Only these second
statistical moments are needed to construct the likelihood 
function for Gaussian theories.  
Let $\vTeta$ be the set of parameters we wish
to constrain.  We represent a set of $\Npix$
observed sky temperatures (e.g., a map) by 
a {\em data vector}, $\vd$, with elements
$d_i \equiv \Delta_b(\hat{n}_i)$.
If the noise is also Gaussian, then the data has a 
multivariate Gaussian distribution, and the likelihood
function is
\begin{equation}\label{eq:like1}
{\cal L}(\vTeta) \equiv {\rm Prob}(\vd|\vTeta)
        = \frac{1}{(2\pi)^{\Npix/2} |\mC|^{1/2}} e^{-\frac{1}{2}
        \vd^t \cdot \mC^{-1} \cdot \vd}
\end{equation}     
The first equality reminds us that the likelihood function
is the probability (density) of obtaining the data vector given a 
set of parameters.  In this
expression, $\mC$ is the pixel--pixel covariance matrix:
\begin{equation}\label{eq:covmat}
C_{ij} \equiv <d_id_j>_{ens} = T_{ij} + N_{ij} = C_{b}(\theta_{ij}) + N_{ij}
\end{equation}     
where the expectation value is understood to be over the
theoretical ensemble of all possible universes realizable with
the same parameter vector. The second equality separates the theory 
covariance matrix ($\mT$) from the noise covariance matrix ($\mN$). One
obtains the third equality from Eq. (\ref{eq:Cbeam}). 
We see that the model (or the
cosmological parameters, $\vTeta$) enters  the likelihood through the 
dependence of $\mT$  on $C_l$ (or $C_l[\vTeta]$).

Depending on the observational strategy, one may have 
either a true temperature map (e.g., COBE), or a set temperature
differences (e.g., MAX).  One could also imagine working 
with more complicated linear combinations of sky temperatures;
this is useful, for example, to customize bands
in Fourier space for reporting power estimates.  Let $\mA$ be
the transformation matrix defining such a 
linear combination of sky temperatures.  Eq.
(\ref{eq:Cbeam}) is accordingly transformed as 
\begin{equation}\label{eq:defW}
T^\prime_{ij} = A_{im}A_{jn}T_{mn} = \frac{1}{4\pi}
        \sum_l (2l+1) C_l W_{ij}(l)
\end{equation}  
where  $W_{ij}(l) \equiv A_{im}A_{jn}P_l(\mu_{mn})|B_l|^2$ is the 
{\em window matrix}.  
The diagonal element, $W_{ii}$, is normally given as the 
window function defining the band in
Fourier space over which the measured power is reported.
In order to estimate this power, one inserts a spectral form
into (9) and finds its normalization as the maximum of the
likelihood function.
For example, the commonly used
{\em flat} band--power, $\delta T_{fb}$, actually represents the 
equivalent logarithmic power integrated over the band:
\begin{equation}\label{eq:dtfb}
C_\ell \equiv 2 \pi [\delta T_{fb}^2 / (\ell (\ell +1)]
\end{equation}
These are the numbers used to 
construct the familiar plot shown in Figure 1.

\subsection{MAX}

The analysis of the MAX experiments is a good example that
helps to clarify the use of 
the window matrix. This experiment re-groups several years of observations 
towards different directions on the sky 
(see Clapp et al. 1994, Tanaka et al. 1996, Ganga et al. 1998 
for details).  We
will analyze the HR, ID and SH campaigns and  
work in pixel--pixel  space.  For example,
the 21 pixels of the MAX ID observations are
described by a $21\times 21$ correlation matrix.  
These pixels are actually differences on the sky, 
defined by the observing strategy, 
so we must compute the window matrix given in Eq. (\ref{eq:defW}).
Consider first the general strategy of a simple two--point difference:
$ d_{diff}  \equiv \Delta_b(\hat{n}_1)-\Delta_b(\hat{n}_2)$, whose variance
is given by 
\begin{eqnarray}\label{eq:Wdiff}
\nonumber
<d_{diff}^2>_{ens} & = & 2[C_b(0) - C_b(\theta)] \\
 & = & \frac{1}{4\pi} \sum_l (2l+1) C_l
        \left\{ 2 |B_l|^2 \left[1 - P_l(\mu)\right] \right\}
\end{eqnarray} 
where the second equality uses Eq. (\ref{eq:Cbeam}).
The window function would be identified as the expression in the curly
brackets. The {\em off--diagonal} terms of $\mW$ 
will depend not only on the
distance between pixels, but also on their relative orientation,
the angular symmetry now being broken by the nature of the
difference.  This means that these terms are
not necessarily expressible as Legendre series.

     In reality, the MAX observational strategy is a sine--wave  
difference, which means that the sky temperature along a scan
are weighted by a sine function, and not just the difference between
two points on the sky.  The window matrix element $W_{ij}(l)$ for two such 
sine--differenced pixels, separated along their common scan axis
by an angle $\Phi_{ij}$ on the sky, may be written as
\begin{eqnarray}\label{eq:Wsin}
\nonumber
W_{ij}(l) = N^2\ B_\ell^2 \sum_r^\ell 
        \frac{(2 \ell- 2 r)! 2r!}{[2^\ell r ! (l-r)!]^2} 
        L^2_{l-2r}(\alpha_o) \\
        \times \cos((l-2r)\Phi_{ij})
\end{eqnarray} 
where $L_i(\alpha_o)= \pi J_1(i \alpha_o)$, $J_1(x)$ is a Bessel function
of the first kind, N is a normalization factor 
and $\alpha_o$ is half the peak--to--peak chop angle
(see White \& Srednicki 1995).  
Note that this is in fact not a Legendre series.

Once all the $W_{ij}$ are calculated, we can construct the theory 
correlation matrix for a given model (defined by either 
an Inflation--generated spectrum, or a set of band--powers)
and compute the likelihood as given in Eq. (\ref{eq:like1}).
This also requires specification of the noise covariance 
matrix, $\mN$, which for MAX we take to be diagonal
with elements equal to the published noise variances.
The MAX observations correspond to 4 well--separated fields 
on the sky; there are therefore no correlations 
between each set (ID, HR, SH, PH), and the 
overall likelihood is just the product of each 
individual field likelihood.  
Results of such an analysis are 
presented at the end of this section and are shown as 
flat band--powers in Figure 1 as the filled diamonds.

\subsection{SASKATOON}

The Saskatoon experiment is described in Netterfield et al. (1997).
Like MAX, the Saskatoon pixels are in reality sky temperature 
differences, although Saskatoon consists of 
a single field observed with different differencing strategies
to probe a variety of angular scales.
The data consist of 39 sets of pixels, each related to
a particular frequency (Ka--31GHz, Q--41GHz), for a particular year
 (1993,1994,1995) and
for a particular strategy; for example, there are 48 pixels 
for the 1994, Q-band, 6--point difference.
As was the case for MAX, these differences are actually the weighted sum
of sky temperatures taken along a scan.  Once the exact weighting
is known, it is straightforward to calculate the window Matrix, 
$\mW$, for each pixel set.  This time, however, there are correlations 
between the different pixel sets, because they look at the
same part of the sky and some sets probe similar scales.
Netterfield et al. proposed the construction of 5 separate bins of
pixels, grouped according to scale such 
that the correlations between bins fall below 20\%. 
Clearly, correlations are larger between the 3--point 
and 4--point differences than between the 3--point 
and 18--point differences.

     We have not considered the data subset corresponding
to the RING observations in the present analysis, but 
instead grouped the CAP pixel sets into 5 bins 
in the same manner as Netterfield et al.  
Finding the power in one
such bin requires calculation of the correlation matrix for
all the constituent pixels.  As an example, the fourth bin takes into 
account the 13, 14, 15--point differences of Q-band 1995, with 
96 pixels for each difference.  The likelihood for the power over this bin,
ignoring correlations, would be the product of 
three individual likelihoods, each involving a $96\times 96$ 
correlation matrix;
but properly taking the correlations into account 
in fact requires a $288\times 288$ correlation matrix. 
The five bins we considered contain, respectively,  $720\times 720$, 
$864\times 864$, $288\times 288$, $288\times 288$ and $384\times 384$ matrices.
We neglect the residual correlations between bins, and use the noise correlation matrix as 
given by Netterfield et al.  This leaves us with 5 likelihoods for each model, 
one for each bin concentrated on 5 different scales in the power spectrum. 
The results are presented at the end of this section, and shown
in Figure 1 as the filled circles.  Our results are in agreement with
those given by Netterfield et al. for the same data ensemble.

\subsection{COBE}

Four years of observations by COBE have resulted in several full--sky 
maps (different wavelength maps and combined maps to reduce Galactic 
emission).  One might expect that Eq. (\ref{eq:like1}) 
and the definition of $\mT$ given in (\ref{eq:Cbeam}) could be directly 
used to compute the likelihood function, since we are dealing
with sky temperatures instead of differences (all referenced to the 
map mean, which is set to zero). 
Note, however, that the likelihood computation requires
the inversion of $\mT$ and a calculation of its determinant,
scaling as $\Npix^3$.
Full--sky COBE maps contain 6144 pixels. The first obstacle to direct 
application of Eq. (\ref{eq:like1}) is the time consuming nature of the 
matrix calculations. 
A second is due to the fact that one must use {\em cut} sky maps 
which remove the Galactic plane.  For the COBE ``custom cut'',
this reduces the number of usable pixels to 3881, 
and a corresponding $3881\times
3881$ correlation matrix to invert.  Now, from 
Eq. (\ref{eq:alm}) we see that
the spherical harmonic coefficients ($a_{lm}$) are independent 
random variables for which the correlation matrix $\mT$ is diagonal,
and thus easy to invert.  
This tempts one to try a Fourier analysis of the COBE maps, but
the sky cut compromises such analysis -- one does not have access to
the actual $a_{lm}$.  One could all the same work with coefficients
calculated only over the cut sky, $\tilde{a}_{lm}\equiv \int d\Omega
\Delta(\hn) Y^*_{lm}(\hn)$, where the integral covers only the cut sky,
but these {\em do not} have the same convenient properties as the
real $a_{lm}$ -- most notably a diagonal correlation matrix.
It is obvious that the sky mask, seen as a convolution in
Fourier space, mixes different $l$ and correlates the
$\tilde{a}_{lm}$.  These $\tilde{a}_{lm}$, and in fact any
coefficients defined on the cut sky, are just another
example of power bands imposed by restricted sky coverage, 
as discussed above, but complicated here by
the fact that they may not be compact over a contiguous
band of multipoles.  

        If one nevertheless wishes to perform
a kind of Fourier analysis on a cut map, then it would be
better (at the very least for numerical stability) to
work with orthogonal functions.  This will not, as just
emphasized, yield uncorrelated quantities 
(orthogonality of the basis functions is not equivalent to
independence over the theoretical ensemble defining the correlation
matrix).   Gorsky (1994) proposed
an elegant method for constructing orthonormal basis functions 
over an incomplete sky map, and we have
used his approach for a likelihood analysis of the ``dcmb'' COBE
DMR sky map. 
Details of the technique can be found in Gorski (1994); here we 
just briefly review the approach.

        Consider an harmonic decomposition as given in (\ref{eq:alm}), 
up to order $l_{max}$.  We arrange our spherical harmonics $Y_{lm}(p)$
(we choose to use real harmonics)  
in a $\Npix$ by $(l_{max}+1)^2$ matrix $\mY$ with general element $(p,\alpha)$ 
equal to $Y_\alpha(p)$, where the index $\alpha\equiv l^2+l+1+m$; in the 
following Latin indices will refer to pixel number, while Greek
indices identify the harmonic function, i.e., $(l,m)$.  
Any function on the sphere may be viewed as a pixel--space column vector
$\vf$ (listing the function values for each pixel, e.g., the data
vector $\vd$), and its harmonic
decomposition is expressible as 
$\vf = \mY\cdot\va$,
where $\va$ is a frequency--space column vector containing the 
harmonic coefficients (the $a_{lm}$).  The matrix $\mY$ lives
in both spaces and transforms an object from one representation to
the other.  
Orthogonality of the $Y_{lm}$ over the full sky and its loss 
over the cut sky are represented by the following relations:
\begin{eqnarray}\label{eq:ident}
\nonumber
\Ompix (\mY^T\cdot\mY)_{\rm full\ sky}  = \mI \ \ \ \ \ 
\Ompix (\mY^T\cdot\mY)_{\rm cut\ sky}  = \mM
\end{eqnarray}
where $\Ompix$ represents the solid angle subtended by
the pixel elements and $\mM$ is a kind of coupling matrix for the 
spherical harmonics restricted to the cut sky; it approaches 
unity as the size of the cut region goes to zero.  Since this 
matrix is positive definite, it may be Cholesky--decomposed 
into the product of an upper triangular matrix $\mU$ and its transpose:
\begin{equation}\label{eq:defL}
\mM = \mU^T \cdot \mU
\end{equation}
The matrix $\mU$ permits a Gram-Schmidt orthogonalization and the
construction of a new basis over the cut sky.  Setting $\mGam=\mU^{-1}$, we 
obtain the new basis functions $\Psi_\alpha(p)$ from  
$\mPsi = \mY\cdot \mGam$.  Their orthonormality is easily
verified: $\Ompix \mPsi^T\cdot\mPsi = \mI$ (notice that 
the $\Psi$ are defined only on the cut sky, so we do not need
to specify this explicitly in the matrix product).
Each new basis function $\Psi_\alpha(p)$ is {\em a linear combination
of spherical harmonics $Y_{\alpha^\prime}(p)$ of 
{\bf lower or equal order, $\alpha^\prime \le \alpha$}}.  
A basis function $\Psi_\alpha$ thus does not correspond to a pure, single 
spherical frequency -- power is aliased in, {\em but only from lower
frequencies}.  The fact that this power leak is only ``red--ward'' is
important, because it preserves a progression towards higher 
frequencies with increasing $\alpha$.
  
     We have implemented this method with $l_{max}=30$ and a total of
961 $\Psi$ functions.  There is very little power
in the COBE maps beyond $l=30$ due to the beam cutoff.
We first decompose the \emph{pixel vector} $\vd$ on the new basis
defined over the cut sky:
\begin{equation}\label{eq:defL}
\vtc = \Ompix\mPsi^T\cdot\vd
\end{equation}
The relation between these coefficients $\tilde{c}$ and 
the $a_{lm}$'s is given by $\vtc = \mU\cdot\va$. 
Written in the spherical harmonic basis, the theoretical 
correlation matrix $\mT$ of Eq. (\ref{eq:covmat}) would be diagonal:
$\mT =  \bra~\va~\cdot~\va^T~\ket = diag(~\bra~a^2_{\alpha[lm]}~\ket~) 
= diag\{C_{l[\alpha]}B_l^2\}$, where $B_l$ accounts for beam smoothing.
Transposed to the the cut sky, this becomes:
\begin{eqnarray}\label{eq:newT}
\mtT = \bra \vtc \cdot \vtc^T \ket = \mU \cdot \bra \va \cdot 
        \va^T \ket \cdot \mU^T \\
\nonumber
 =  \mU \cdot diag\{C_l \cdot B_l^2\} \cdot \mU^T
\end{eqnarray}
This matrix has no a priori reason to be diagonal. 

        The noise correlation matrix 
must also be projected from the original pixel space into
the observational space defined by the new basis:
$\mtN = \Ompix^2~\mPsi^T~\cdot<~\vd~\cdot~\vd^T~>~\cdot~\Psi$
Since the noise is essentially uncorrelated in the COBE maps,
the pixel--pixel noise correlation matrix is diagonal 
(but not proportional to the identity matrix!) and the
projected noise correlation matrix reduces to
$(\mtN)_{\alpha\alpha^\prime} = \Omega_{pix} 
\sum_p \sigma^2_N(p) \ \Psi_\alpha(p) \cdot \Psi_{\alpha^\prime}(p)$,
where $\sigma_N(p)$ is the noise variance at pixel $p$.

        We may now rewrite the likelihood function   
of Eq. (\ref{eq:like1}) in the new basis:
\begin{equation}\label{eq:like2}
{\cal L}(\vTeta) \equiv {\rm Prob}(\vtc|\vTeta) \ 
        \propto \ \frac{1}{ |\mtT + \mtN|^{1/2}} e^{-\frac{1}{2}
        \vtc^t \cdot (\mtT + \mtN)^{-1} \cdot \vtc}
\end{equation}
We computed the likelihood of a suite of cosmological
models with this function, and   
our results are summarized in the figures at the end of this section.
Note, however, that the points plotted in Figure 1 do not issue from
our analysis, but come rather from Tegmark \& Hamilton (1997) and
are based on the Fisher matrix, as described in detail by
these authors.

\subsection{Likelihood Results}

        We now detail some constraints obtained from our full 
likelihood analysis of the COBE, MAX and Saskatoon data
sets, following the methods outlined above.  
These results are interesting in their
own right, and they will also serve as a benchmark against which
other methods will be subsequently evaluated.  {\em It is only this
complete analysis which permits us to perform an in--depth evaluation
of alternate methods attempting to approximate the full likelihood
approach}.  It being computationally impossible to explore
a large parameter space, we shall illustrate with a series of open
Inflationary models, varying $\Omega$, 
$H_o$ and $Q$ in the respective ranges $[0.1, 1, \mbox{\rm step}=0.1], 
[15, 100, \mbox{\rm step}=5$km/s/Mpc$], [11,23, \mbox{\rm step}=1 \mu$K$]$; 
the spectral index $n$ is fixed to $1$, $\Omega_b h^2 = 0.018$ 
(Olive, Steigman \& Walker 2000; Tytler et al. 2000) and 
$\Omega_\Lambda=0$. We consider neither reionization, nor 
the presence of gravitational waves.  This leaves us with 2340 models
whose likelihood were computed.  As mentioned, the independence of
the selected experiments implies that the total likelihood function
is simply the product of each individual likelihood.

     The best--fit model among this set corresponds to 
the parameter values $\Omega=1$, $\Ho=30$~km/s/Mpc, $Q=18 \mu$K.  
We note that this is in many ways a toy model, being 
based on a restricted data and parameter set.  The zero
curvature does agree with recent CMB results, such
as BOOMERanG and MAXIMA--1 (Jaffe et al. 2000).
Since we do not include a cosmological constant in 
our analysis, this implies a critical universe, 
which does not satisfy other cosmological constraints,
for instance those arising from SNIa distance measurements
(Riess et al. 1998; Perlmutter et al. 1999) and cluster
evolution (Bahcall et al. 1999) ; although it 
remains consistent with some analyzes of cluster evolution
(Blanchard et al. 2000).
Nevertheless, this toy model is sufficient for our primary 
aim of testing analysis methods.
We first discuss the
GoF of this model, finding that it is acceptable (or that
the fluctuations are consistent with a Gaussian origin), and then
present the parameter constraints.

\subsubsection{Goodness of Fit}

The ability to compute a GoF is an essential part of parameter 
estimation.  Given a set of data points and models, one can 
always find a 
``best model'' and construct many different 
ways of giving confidence intervals on the parameters.
The essential question we must answer is double: 
does the best model actually reproduce 
the data well? and if so, what are the confidence intervals 
defined around the best--fit model.  
The GoF attempts to quantify the first part of the question, while
contours of our likelihood surfaces respond to the second.   
If the best model does not pass the GoF, then there is no 
point in defining confidence intervals -- the suite of chosen
models is ruled out.  

     Maximization of the 
likelihood gives us our best estimate for the parameters, 
$\vTheta_{\rm best}$.  To test the GoF, we use the following
statistic
\begin{equation}\label{eq:defG}
G = \sum_{i=1}^{N_{\rm bin}} \vd^{(i)t} \cdot\ \mC^{(i)} (\vTheta_{\rm best}) 
    \cdot \ \vd^{(i)}
\end{equation}
where the sum is over all experiments (COBE, MAX, Saskatoon) and 
relevant bins, as discussed above. The quantity 
$\mC^{(i)}(\vTheta_{\rm best})$ is the correlation matrix
evaluated for the best--fit model.
As might be wished, the quantity $G$ is distributed like a $\chi^2$ 
with a number of degrees--of--freedom equal to the total
number of pixels summed over all experiments and bins: $\Npix=2521$.
This means that if our ``best model'' is a good fit, 
then the value of $G/\Npix$ should be near 1.
For our best model ($\Omega=1$, $\Ho=30$ km/s/Mpc$^2$, $Q=18 \mu$K),
we find $G/\Npix = 1.0473$\footnote{In comparison, a model with  
$\Omega=0.3$, $\Ho=60$ km/s/Mpc$^2$ and
$Q=18 \mu$K gives $G/\Npix=1.23$ with 
$\Npix=2521$.}.  Thus, according 
to this test the model
does indeed provide an adequate fit to the data, and so we may 
now move on to consider parameter constraints.  This also implies
that the data are consistent with a Gaussian origin for the
temperature fluctuations, at least according to this 
particular statistic.

\subsubsection{Constraints}

\begin{figure}\label{fig_powplot}
\begin{center}
\resizebox{\hsize}{!}{\includegraphics[angle=0,totalheight=8.4cm,
        width=8.cm]{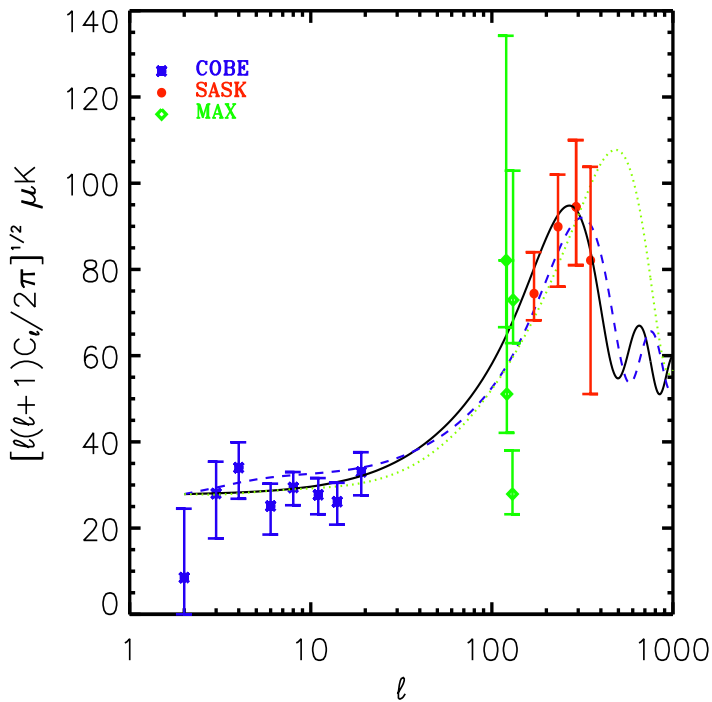}}
\end{center}
\caption{CMB power spectrum estimates.  Flat band--powers
are shown as a function of multipole order $l$.  The data correspond to our
estimates for MAX \& Saskatoon. The COBE band--powers were
obtained by Tegmark \& Hamilton (1997), and are not based on 
a likelihood analysis.  The solid (black) line is the best fit
from the full likelihood analysis; the dashed (blue) line is the 
best--fit $\chi^2$ model; and the dotted (green) line is for
$\Omega=0.5$ and $\Ho=25$~km/s/Mpc (see text).}
\end{figure}

     For each experiment or bin, our results take the form 
of a three--dimensional matrix providing the likelihood 
over the parameter space.  Constraints are
presented by projecting this matrix onto 2D parameter 
planes as follows:  we construct the surface
$LL(p_1,p_2)\equiv - 2\log {\cal L}$ over the 2D plane
of interest $(p_1,p_2)$ by either fixing the third parameter,
or by letting it take on the value that minimizes
$LL(p_1,p_2)$ (``marginalizing'').  Contours of equal 
confidence are then
defined by adding specific values $\Delta$
to the minimum of the surface $LL(p_1,p_2)$,
e.g., $\Delta=1,4$.
If the likelihood were Gaussian (which it
is not), then these particular values would identify, 
respectively, the $68.3 \%$  and $95.4 \%$ confidence 
limits of $p_1$ or $p_2$ when projected onto these
axes.

Figure 2 shows the constraints over the 
$(\Omo,\Ho)$--plane with $Q$ marginalized.
Inflationary power spectra are characterized 
by a succession of oscillating peaks (``Doppler peaks'')
with the first around $\ell \sim 200$.
These are exactly the scales to which Saskatoon is sensitive. 
The position of this first peak is strongly related to the 
curvature of the universe
($\Omk = 1 - \Omb - \Omc - \OmL$),
and therefore to $\Omo=\Omb+\Omc$, 
given that we have set $\OmL = 0$. 
It is then not surprising that we have a good constraint on
the position of the peak, and so on $\Omo$. 
Combined with COBE, we are able further to fix the 
height of the first peak relative to the large--scale plateau
of the Sachs--Wolfe effect.  This height is controlled by 
the quantity $\Omega h^2$, providing an additional
constraint of these parameters.

It is noteworthy that with just these three experiments, 
albeit well selected, we obtain nontrivial constraints on 
the two free parameters $\Omega$ and $H_o$, at least within
the chosen open Inflationary context.  The robust conclusion,
clearly applicable beyond the present restricted context, is 
that large curvature is disfavored by the observed position
of the peak, a conclusion 
reached by many authors previously (Lineweaver et al. 1997, Bartlett et al.
1998ab, Bond \& Jaffe 1998, Efstathiou et al. 1999, 
Hancock et al. 1998, Lahav \& Bridle 1998, 
Lineweaver \& Barbosa 1998ab, Lineweaver 1998, Webster et al. 1998,
Lasenby et al. 1998, Dodelson \& Knox 1999;
Tegmark \& Zaldarriaga 2000; Knox \& Page 2000),
and confirmed by the recent BOOMERanG and MAXIMA--1 
results (de Bernardis et al. 2000; Hanany et al. 2000;
Jaffe et al. 2000). 

     These results obtained
from a full likelihood analysis now permit us to 
test other, less time--consuming methods that
have been proposed as an approximation to such
a complete treatment.

\begin{figure}\label{fig_saskmaxcob}
\begin{center}
\resizebox{\hsize}{!}{\includegraphics[angle=0,totalheight=9cm,
        width=8.9cm]{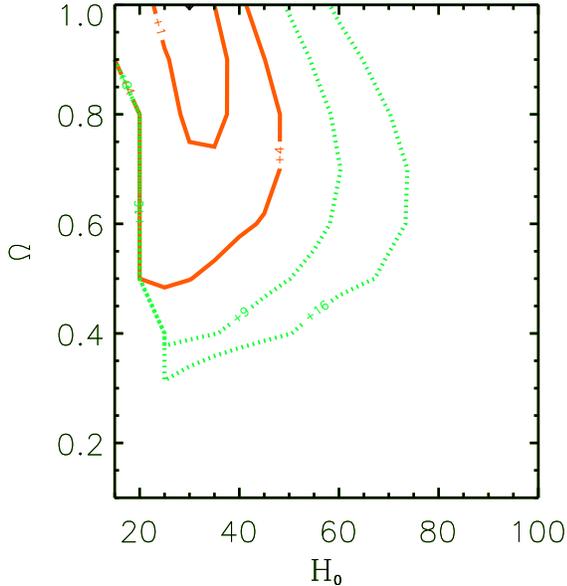}}
\end{center}
\caption{Likelihood contours for the combined analysis of COBE, 
Saskatoon and MAX, with Q marginalized.  Solid (red) contours are
for $\Delta = 1, 4$, and dotted (green) contours are for 
$\Delta = 9, 16$}
\end{figure}


\section{Radical Compression -- the Power Plane}

     The time--costly nature of a full likelihood analysis
over pixel space prevents its general application to
actual CMB data sets (Bond et al. 1998, 2000; Borrill 1999ab;
Kogut 1999).  This has motivated the development of 
faster, but approximate, methods.  These are 
{\em approximate} in the sense that they do not 
necessarily retain all relevant experimental information, 
as does the full likelihood treatment.  It is therefore
important to compare the results of these approximate 
methods to those of a complete likelihood treatment.  
Because we have performed a
complete likelihood analysis, we are now in a position to do 
just this.

     All currently proposed approximate approaches 
use power estimates as their starting point.
As discussed in the introduction, the power spectrum
may be considered as a compressed form of the data,
because there are many fewer independent power 
points than original pixels\footnote{This 
compression is a consequence of the 
assumed statistical isotropy of the fluctuations that
demands that the correlation function only depend on
angular separation -- there are many fewer
separations than original pixels on the sphere.}. 
Gaussian fluctuations observed in the absence of noise
over the full sky are completely described
by their set of multipole power estimates
$C_l$.  Real observations, however, contain noise and cover 
only limited amounts of sky.  It is then no longer possible to
uniquely decompose the sky signal;  the noise and limited 
sky coverage (equivalent to a window function) reduce the 
spectral resolution and correlate power estimates.
Flat band--power estimates, as shown in Figure 1, represent
a particular attempt to express an experimental result
in the power plane under such circumstances.  
There is no guarantee that the reduction to 
a set of flat--band powers does not involve
the loss of pertinent information, i.e., that
it is a kind of {\em lossy} data compression.  
This raises an important question concerning 
the adequacy of any method based on flat--band estimates
to reproduce the complete likelihood results 
(which we take to be the defining goal of any 
proposed analysis scheme).  
Once given a power estimate, one must then decide
how to use it in a correct statistical analysis; different
choices lead to alternative approximate methods.  

     We begin this section by first examining the 
accuracy (always compared to our complete likelihood
results) of different ways of using flat--band 
estimates. 
In the second part of the section, we return to the
fundamental question raised in the previous paragraph
and study in greater detail the whole premise of the 
flat band--power approach -- namely, whether  
any of these methods are able to 
recover all the relevant information contained in the 
likelihood analysis.  We will discover shortcomings
of the flat--band approach that will lead us to propose 
better approximate methods.


\subsection{$\chi^2$ minimization}

The most obvious way of finding ``the best model'' given 
a set of points and errors is the traditional $\chi^2$--minimization. 
This would appear to apply to our situation: we are given the points 
plotted in Figure 1, $\delta T^{obs}(N)$, with errors,
and we have a large set of models depending on diverse 
parameters that we can express in terms of temperature 
fluctuations, $\delta T^{model}$. We would therefore 
just minimize
\begin{equation}\label{eq:chi21}
\chi^2(\vTheta)=\sum^{N_{exp}}_{N=1}\left[\frac{\delta 
    T^{obs}_{\ell_{eff}}(N) -\delta T^{modele}_{\ell_{eff}}(N,\vTheta)}
  {\sigma_N}\right]^{2}
\end{equation}
This is the most basic and obvious way of estimating the parameters,
$\vTheta$.  Improvements may be added to take into account the 
asymmetry of the error bars and the effects of the window 
function.  For the former, one would take a different $\sigma$ in 
Eq. (\ref{eq:chi21}) if the model prediction is greater 
($\sigma +$) or lower ($\sigma -$) than the observed value,
and the model predictions may simply be convolved with the window 
function.

The main problem with this approach is that it 
treats the flat band--power estimates as Gaussian 
distributed data.  As we saw at the beginning of the 
first section, Gaussian temperature
fluctuations, such as occur in Inflationary models, 
lead to pixels that are Gaussian random distributed, 
and so also the $\alms$. Since band powers measure 
the variance of these random sky fluctuations, 
the distribution of a power estimate {\bf
cannot} be Gaussian.
It is something more akin to
a $\chi^2$ distribution.  While it is of course
true that this tends to a Gaussian  
as the number of independent 
values (pixels) entering the power estimate
increases, this does not always apply in practice
to CMB data: consider MAX as an extreme example
with only 21 pixels (and not effectively
independent due to correlations).  It is not
true that the relevant number of degrees--of--freedom
corresponds to the multipole order to which the experiment
is sensitive; it is rather the number of effectively
independent pixels, and any experiment
will always be limited by a small number of 
pixels over the largest scales probed.

        There is a perhaps less well--appreciated aspect
of this issue.  If one wishes 
to reproduce as closely as possible the
complete likelihood analysis outlined
in the previous sections, then what one really needs to know
is ${\cal L}(params) = {\cal L}[\delta T_{fb}(params)]$,
which is not the same thing as the distribution
of the power estimator\footnote{Note that one 
must distinguish the likelihood function for 
the power from the the distribution of the
power estimator (the maximum of the likelihood 
function).  Neither is Gaussian.}.  
To be precise, the power estimator is given
by the maximum of the band--power likelihood function.
Its distribution can be determined by a Monte
Carlo for an adopted underlying model
(set of parameters).
This remark is particularly relevant considering
the nature of the error bars given in plots
such as Figure 1: these are confidence intervals
based on the likelihood function, and not
the second moment of the power estimator
distribution.  
Due to its simplicity, the
$\chi^2$ approach lies at the basis of most
current efforts to constrain cosmological parameters with
CMB data, despite these shortcomings.  It is thus of some importance to 
test its (a priori dubious) adequacy.

\begin{figure}[t]\label{f:LSM18}
\begin{center}
\resizebox{\hsize}{!}{\includegraphics[angle=0,totalheight=8cm,
        width=7.9cm]{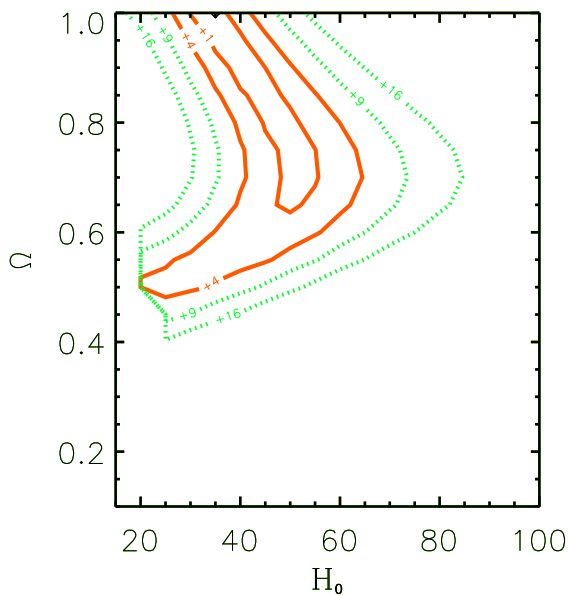}}
\end{center}
\caption{Likelihood contours for the combined analysis of Saskatoon and
MAX with $Q=18 \mu$K. Contours have the same definition as in Fig. 2.
The best--fit model is indicated by the diamond at the top of the Figure --
$\Omo=1$ and $\Ho=35$ km/s/Mpc.}
\end{figure}

     Results from $\chi^2$--minimization 
will be compared to our complete likelihood analyzes.
For clarity, and to emphasize pertinent aspects of 
the problem, we continue to work in a restricted 
framework by fixing the normalization $Q=18~\mu$K 
and combining only the MAX and Saskatoon experiments 
(7 flat band--powers).  
\begin{figure}\label{f:CSM18}
\begin{center}
\resizebox{\hsize}{!}{\includegraphics[angle=0,totalheight=8cm,
        width=7.9cm]{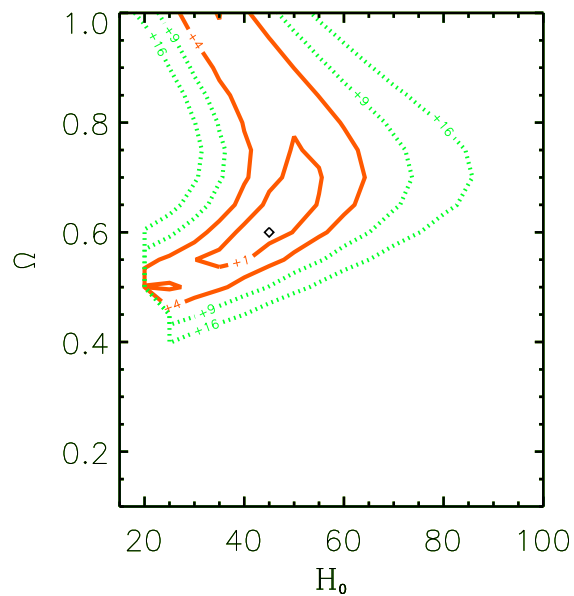}}
\end{center}
\caption{$\chi^2$ contours for the combined analysis of Saskatoon and
MAX with $Q = 18 \mu $K. Contours have the same definition as in Fig. 2.
The best--fit model is again indicated by the diamond and corresponds
to $\Omo=0.6$ and $\Ho=45$ km/s/Mpc.}
\end{figure}
Figures 3 and 4 show constraints in this restricted context for, 
respectively, the likelihood analysis and the $\chi^2$ method.
We clearly see a difference in the inner contours between the two
approaches; and the ``best model'' also changes radically depending 
on the method.  To see if the prefered models agree with our intuitive 
``$\chi$--by--eye'', we plot them in the power plane of Figure 1.
They appear both to agree perfectly with Saskatoon, while 
trying to pass between the MAX points.
 
The largest contribution to the $\chi^2$ comes from MAX HR (lowest
MAX point in Figure 1). 
It is natural to ask if this outlier alone can account for the 
difference between the two methods.  A simple way to test this 
is by substituting just this experiment's true likelihood 
by its $\chi^2$ in an otherwise full likelihood analysis.  This
results in the contours of Figure 5, which are to be compared 
to those of Figure 4.
\begin{figure}\label{f:lcSM18}
\begin{center}
\resizebox{\hsize}{!}{\includegraphics[angle=0,totalheight=8cm,
        width=7.9cm]{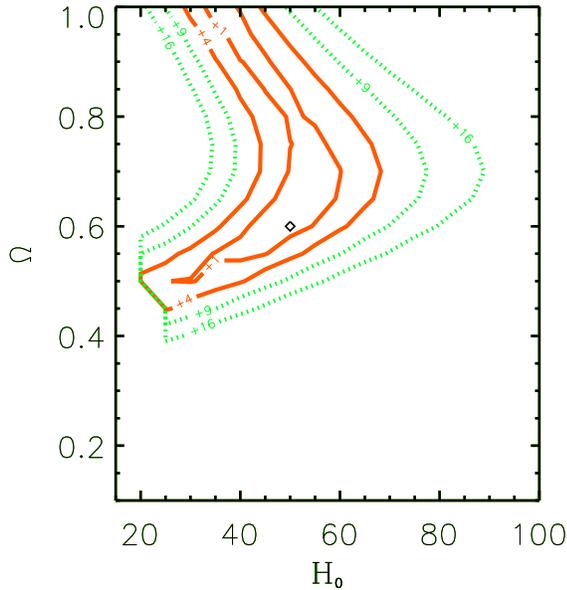}}
\end{center}
\caption{Result of combining the likelihood for each bin with the 
Gaussian for MAX HR ($\chi^2$).  The contours close around a 
best--fit model similar to that of Figure 4, demonstrating the
importance of this single point to the final $\chi^2$ results.}
\end{figure}
The fact that the contours begin to close around $\Omo=0.6$ and 
$\Ho=45$ km/s/Mpc, as in Figure 4,
indicates that indeed this one {\em outlying}
point is capable of radically changing the confidence 
contours and the deduced ``best model''.  Recall the 
supposition of a Gaussian distribution implicitly adopted
by the $\chi^2$ approach.  Here, we have 
precisely a case where the ``best models'' are typically far
into the wing of the distribution function for this outlier, 
and so it is perhaps not too surprising that the $\chi^2$ 
method is at some odds with the complete likelihood analysis.

        To further explore the issue,
consider the difference in terms of the one--dimensional 
flat band--power likelihood function.  Figure 6 shows the 
distributions used by the two methods. 
The solid (black) curve is the true likelihood function while the 
dashed--3--dotted (red) 
curve shows a two--tailed Gaussian, as employed with 
Eq. (\ref{eq:chi21}).  The two distributions clearly diverge 
for $\delta T \ket 50 \mu K$ ($LL$ greater 
than 6), exactly where the two models of Figure 1 pass.
From this figure we see that a model with a temperature 
fluctuation of $60 ~\mu$K falls on the 99\% confidence boundary 
($\Delta = 9$) in the
likelihood analysis, but at 99.99\% confidence
($\Delta = 16$) in the $\chi^2$--minimization.  This 
makes a significant difference to the final contours.
All the same, we note that the final 95\% confidence contours 
are essentially the same for the two methods.

\begin{figure}\label{f:func}
\begin{center}
\resizebox{\hsize}{!}{\includegraphics[angle=0,totalheight=8cm,
        width=7.9cm]{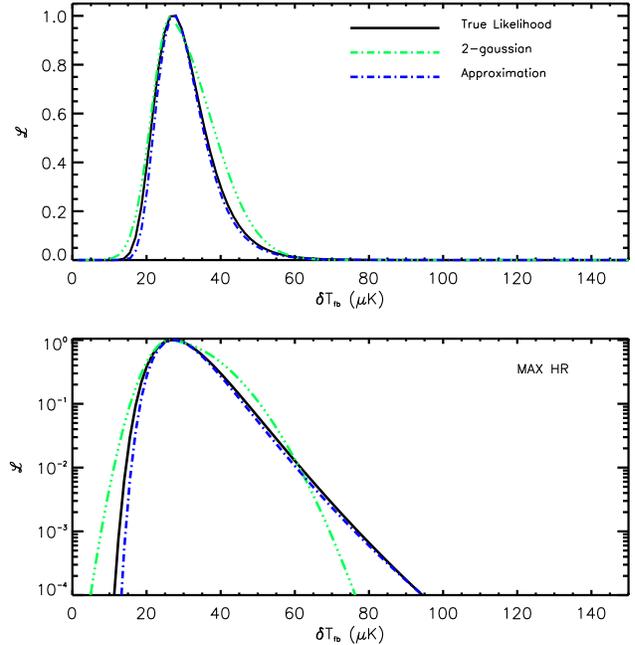}}
\end{center}
\caption{One--dimensional flat--band likelihood curve (solid, black
line) shown together with our approximation (dashed, blue line)
and the Gaussian used in the $\chi^2$ analysis (dashed--3--dotted,
green curve).}
\end{figure}

In summary, the presence of outliers
can significantly change the constraints
of a $\chi^2$--minimization
relative to the true likelihood analysis,
because a Gaussian is a particularly 
bad approximation of the latter in
the wings.  In the above example, 
the overly rapid fall--off of the Gaussian
relative to the true likelihood `pulls'
the contours towards lower $\Omega$ and favors 
an entirely different best fit model
than that selected by the 
likelihood analysis.  On the other hand, the constraints at
the 95\% confidence level are nearly the same.
Thus, it is rather difficult to arrive
at a firm conclusion concerning the accuracy
of the $\chi^2$ method, but 
this example would seem to indicate that some caution
is required when applying the method.

\subsection{Other methods}

How can we improve on a $\chi^2$ minimization while still 
using only flat band--powers? 
Some authors (Bartlett et al. 2000, hereafter
Paper 1; Bond et al. 2000, Wandelt et al. 2000) have 
proposed functional forms (differing from a Gaussian)
to approximate the band--power likelihood.    
In the present section, we study in detail the performance of 
the approximation we developed in Paper 1.  There it was 
shown that the proposed approximation works quite well in
fitting the {\em one---dimensional} flat--band likelihood
function (see Figures 2, 3, 4 in Paper 1 and http://webast.ast.obs-mip.fr/cosmo/CMB).
We may still wonder, however, if the approximation is good 
enough to reproduce the full likelihood constraints  
over the entire ($\Omega, H_o$)--plane.  This is our 
present concern. 

Consider first that the approximation does indeed improve
on the $\chi^2$ minimization, as we demonstrate in Figure 7
by substituting the approximation for the true 
likelihood of MAX HR, in an otherwise full likelihood analysis.
\begin{figure}[t]\label{f:LSM18}
\begin{center}
\resizebox{\hsize}{!}{\includegraphics[angle=0,totalheight=8cm,
        width=7.9cm]{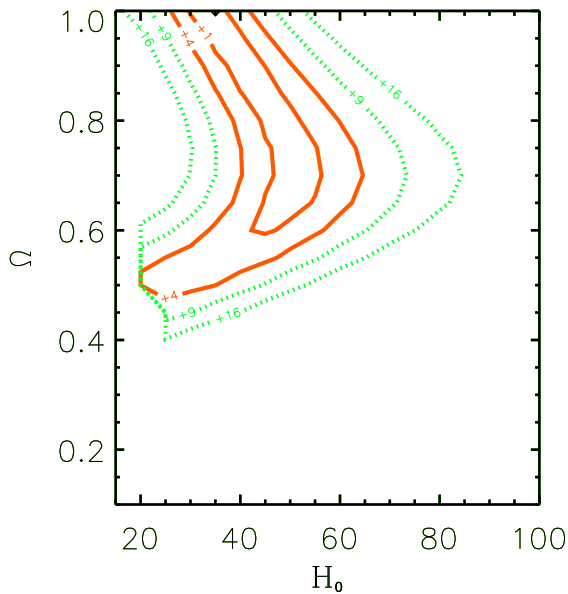}}
\end{center}
\caption{Result of combining the likelihood for each bin with our
approximation for MAX HR. The resulting contours and the best--fit
model better match the full likelihood results of Figure 3 than
the $\chi^2$ results of Figures 4 and 5.}
\end{figure}
With the approximation, we able to recover almost the same contours 
as in Figure 3, thereby eliminating the over--importance
given previously to this outlier by the $\chi^2$ method. 
Further, when we substitute the approximation for all 
the power points, we reproduce the full likelihood 
constraints better than the
$\chi^2$ minimization.  There remain, nevertheless,
some slight differences between the approximation 
and the full likelhood, acting in
the same sense of rejecting the models at high 
$\Omega$ prefered by the likelihood analysis.
We may wonder whether
these remaining differences are due to the fact
that the approximation does not exactly reproduce
the one--dimensional flat--band distribution, or to
something more profound?  In other words, supposing
that we had the {\em exact} 1D flat band--power likelihood
function, could we then recover the full likelihood
contours?   

To examine the issue in depth, we compute the exact 
band--power likelihood function for each experiment/bin 
over a range of $\delta T_{fb}$.  The likelihood of any given 
model is then found by simple interpolation.  
We would like to point out at this point that in fact this technique
would be more accurate and almost as fast as a $\chi^2$--minimization,
if it proves accurate. 
Therefore, we invite people to published the one dimensional
likelihood functions and to make them available for each new (and old if
possible) experimental result.

Figure 8 shows the results, which are to be compared to
the true likelihood constraints of Figure 3.  We see that 
differences persist, including the shifted
best--fit model, even though the 
band--power likelihood has been as well approximated 
as possible.  Thus, the remaining difference in contours
is not due to the use of an approximation to ${\cal L}(\delta T_{fb})$,
but apparently to {\em information lost in the reduction to flat 
band--power estimates}.  In the following discussion
on this point, we will refer to all techniques 
using flat--band estimates, and not pixel 
values, as {\em generalized} $\chi^2$ techniques.

\begin{figure}[t]\label{f:CbSM18}
\begin{center}
\resizebox{\hsize}{!}{\includegraphics[angle=0,totalheight=8cm,
        width=7.9cm]{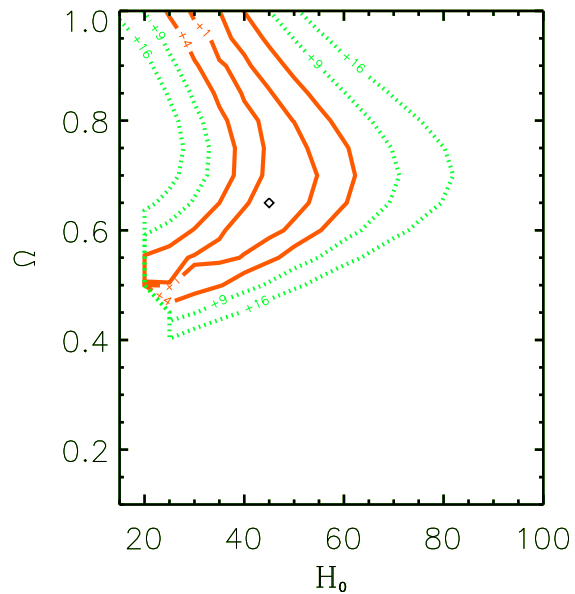}}
\end{center}
\caption{Result of interpolating the exact flat--band likelihood 
for each bin.  Notice that some differences still remain compared
to the complete likelihood analysis.}
\end{figure}

\subsection{Are Band Powers Enough?}

The residual differences just noted bring us
back to the comments made at the beginning
of this section concerning the reduction of 
an observation to a flat band--power 
estimate.  Any given experiment is sensitive
to a number of different angular separations, 
and the totality of its information content is thus
contained in as many numbers.  For example, the set of 
parameters consisting of the {\em distinct} elements 
of the theory covariance matrix $\mT$ provides
a complete description\footnote{For a simple map, these
quantities correspond to estimates of the angular correlation
function.  Not being independent quantities 
in general, one
may then look for a subset of uncorrelated 
parameters, e.g., by diagonalizing the covariance matrix.
Note, however, that one must adopt a priori a model
to define the theory covariance matrix to be diagonalized.
Strictly speaking, the new  parameters are then only
decorrelated for this one model.}.  The likelihood
should then be viewed as a function
of these parameters, each with its own frequency--space
distribution given by the elements of the window
matrix $\mW$.
In this context, we see that the reduction to a 
single flat band--power estimate is but the most 
crude (order zero) representation of an experimental 
result, and we should not be surprised if it is 
not always sufficient to the task.  

     We may attempt to quantify the 
information missing in the 
flat--band representation as follows:
instead of the flat spectrum of Eq. (\ref{eq:dtfb}), 
we employ a spectral form with {\em two} free
parameters -- a normalization $\dTnorm$ and
an effective slope, $m$,  
\begin{equation}\label{eq:modfbm}
\delta T =  \dTnorm (\frac{\ell}{\ell_{eff}})^m
\end{equation}
and treat the likelihood as a 
function of both; the quantity $\ell_{eff}$ is
the usual effective multipole defined over the 
experimental window function.  
If both parameters are constrained 
by the data, then at least these two parameters 
are required to model the likelihood.  
Generalized $\chi^2$ techniques fix the slope to zero
($m\equiv 0$)
and therefore neglect this information.
We proceed by 
first restricting ourselves to the third bin of Saskatoon
(chosen at random) in order to gain some insight.  
We return to the complete data set at the end of the
section, where we propose a new technique that 
accounts for the effects we now describe. 

     Figure 9 shows the constraints imposed by the third Saskatoon 
bin (see Figure 1) in the $(\Omega,\Ho)$--plane for $Q=18\mu K$. 
The black (solid) curves follow from a complete
likelihood analysis, and the green (dot--dashed) ones from an interpolation
of $L(\delta T_{fb})$, as described previously.
\begin{figure}[t]\label{f:toutS4}
\begin{center}
\resizebox{\hsize}{!}{\includegraphics[angle=0,totalheight=8cm,
        width=7.9cm]{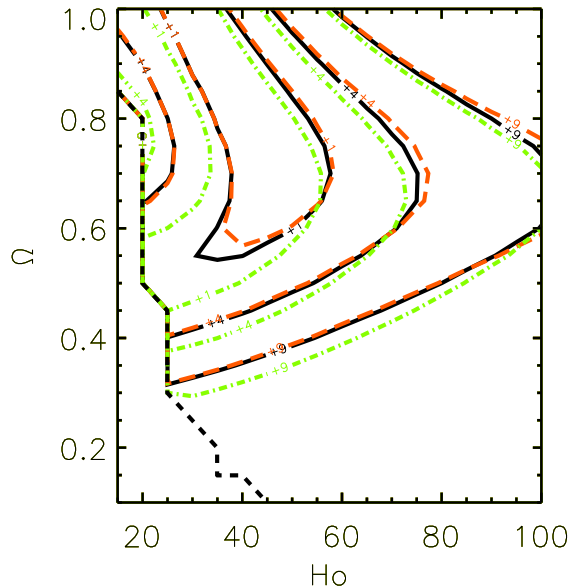}}
\end{center}
\caption{Constraints in the ($\Omega,H_o$)--plane using 
different analysis techniques for
the third bin of Saskatoon and for $Q=18\; \mu$K.
Likelihood contours are in black (solid line), 1D flat--band
likelihood constraints in green
(dot--dashed line) and the interpolation of the likelihood {\em surface} 
over the two parameters $(\dTnorm,m)$ are in red (dashed line).}
\end{figure}
As an illustration, notice that models with 
very low $\Ho$ and intermediate $\Omo$ are more
acceptable to the generalized $\chi^2$, but fall outside 
the inner contour of the \LA\ constraints.
Typically, these models all  
rise through the third bin of Saskatoon, i.e., 
their effective slope is positive and far from zero;
an example is shown as the dotted line in Figure 1.
That the data in fact prefer a {\em negative} slope 
is demonstrated in Figure 10.  Here we show 
the likelihood {\em surface} constructed
over $(\dTnorm,m)$ using Eq. (\ref{eq:modfbm});
we immediately observe that falling spectra
are favored ($m\sim -0.75$) on this scale.
This was also noted by Netterfield et al. (1997),
and the same analysis applied to the other 
Saskatoon bins finds the same trends as noted 
by these authors (even though we do
not include the RING data here). 
The fact that $m$ is at all constrained
implies that the likelihood is sensitive
to the local effective spectral slope.
Neglect of this information by all 
generalized $\chi^2$ methods lies at the
origin of the differences noted between the flat--band
results and the full likelihood in Figure 9.
We will clearly do better by locally approximating
any given spectrum by Eq. (\ref{eq:modfbm}) and 
using the resulting $\dTnorm$ and $m$ to find
the corresponding approximate likelihood for 
the model.  In practice this may be done by
interpolating over a pre--calculated grid
in the $(\dTnorm,m)$--plane, shown for the 
third saskatoon bin in Figure 10.  When applied to 
this bin, we find the red (dashed)
contours in Figure 9, a more faithful 
reconstruction of the full likelihood analysis.  

     We thus propose a new approximate method
taking into account not only in--band power, but
also information on the local shape of the power
spectrum.  It is important to remark that 
{\em it is little more complicated or time consuming than 
approximations based on simple flat--band likelihood curves}; 
the likelihood surface
in 2D may be calculated once, and then interpolated 
for any model.  More generally, one may imagine that
additional parameters would useful when dealing with
higher signal--to--noise data sets, such as expected
with next generation observations.  In reality,
the exact choice of parameters is debatable and
should depend on the number and definition of
spectral bands.

\begin{figure}[t]\label{f:dtfb}
\begin{center}
\resizebox{\hsize}{!}{\includegraphics[angle=0,totalheight=8cm,
        width=7.9cm]{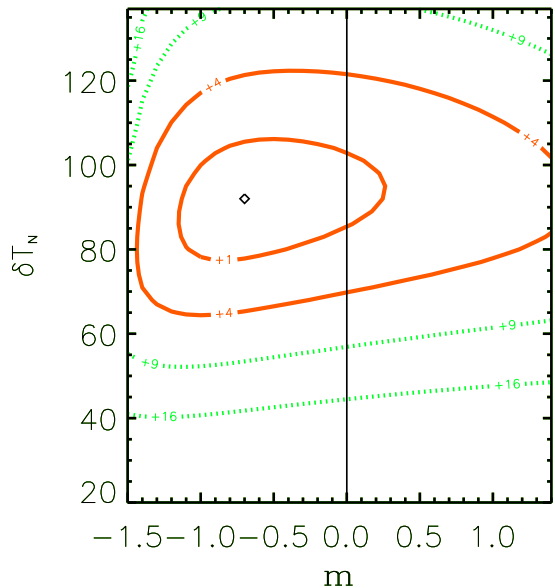}}
\end{center}
\caption{Likelihood contours in the plane ($\dTnorm,m$)--plane 
(i.e., in--band power, in--band slope) 
for the third Saskatoon bin.  Flat band--power estimates 
correspond to the value of $\dTnorm$ that maximizes the
likelihood along the vertical line centered on $m=0$.
}
\end{figure}

\subsection{Discussion}

     The goal of these power--plane methods
is to approximate as closely as possible
the full likelihood analysis, and two
central issues are the nature of the power
estimates and their use in a statistical analysis.
Many current CMB constraints
follow from standard $\chi^2$ techniques using
flat--band power estimates and associated errors.  The 
obvious objection is that neither the 
distribution of this power estimator, nor its
likelihood function is Gaussian (e.g., Figure 6).  
As we have seen from the comparison of Figures 3, 4 and 5, 
the false assumption of Gaussianity on the
part of the $\chi^2$ causes it to 
be overly sensitive to ``outliers'', leading
to a possible bias in the best--fit model 
(as in our examples) and a distortion of 
the confidence intervals.  Other methods
overcome this deficiency by adopting simple,
non--Gaussian analytic functions to 
approximate the likelihood of 
$\fbP$ (Paper 1; Bond et al. 2000; Wandelt et 
al. 2000).  Although performing quite admirably,
and better than the $\chi^2$, the approximate
approach is unable to fully reproduce the
exact likelihood constraints.  
The fact that this remains true even if the {\em exact}
flat--band likelihood curve is used, say
by interpolating over a pre--calculated table of
values (Figures 3 and 8), indicates
that the reduction to flat band--powers 
losses pertinent information.

     In general terms, a set of temperature measurements 
requires as many parameters as unique pixel separations
to be fully described, and this may be as large as or 
smaller than the original number of pixels, depending 
on the exact geometry of the observations.
One implication is simply that, in principle, an
experiment is sensitive to details of the power
spectrum other than just the amplitude averaged over
the range of probed scales; e.g., an additional
sensitivity to the local slope.
Whether this information is in practice useful 
depends principally on the signal--to--noise ratio. 
The critical test is to see if these additional spectral
characteristics are constrained by the likelihood function.
We studied this for the Saskatoon and MAX data sets
and found that in some cases the local
power spectrum slope is in fact constrained, thereby
implying that useful information is lost in the 
reduction to a single band--power.  This
is illustrated in Figure 9 for the third Saskatoon
bin.  This bin actually prefers a slightly falling 
spectrum, and it is interesting to point out that 
the first bin, in contrast, prefers a slightly positive
slope.  The Saskatoon data thus provide additional
discrimination on the first Doppler peak's position
than one would deduce from the band--powers alone. 

        Returning to parameter constraints with this new
insight, we were able to fully reproduce the likelihood
constraints in the $(\Omega,\Ho)$--plane by modeling
each bin likelihood (when needed) as a function of
two spectral parameters -- a local amplitude and slope,
Eq. (\ref{eq:modfbm}).  This is demonstrated in
Figure 9 for the third Saskatoon bin alone, and in
Figure 11 for the ensemble of COBE, MAX and Saskatoon bins.
This procedure is essentially as practical as any
approximation to the 1D flat--band likelihood function:
one only needs to calculate the likelihood {\em surface}
over $\dTnorm$ and $m$ once and use the tabulated
values as a basis for interpolation to any other 
parameter values.  
 
\begin{figure}[t]\label{f:touttout}
\begin{center}
\resizebox{\hsize}{!}{\includegraphics[angle=0,totalheight=8cm,
        width=7.9cm]{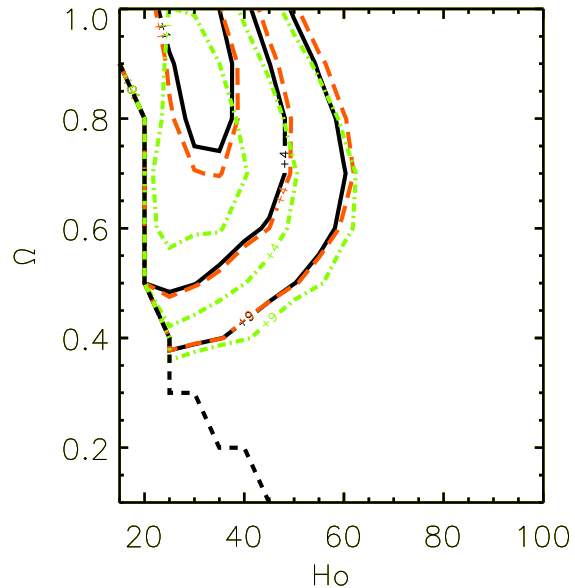}}
\end{center}
\caption{Contours in the ($\Omo,H_o$) plane based on interpolation 
of the likelihood {\em surfaces} $(\dTnorm,m)$ (see text) for
the entire set of Saskatoon and MAX bins, and the true likelihood analysis 
for COBE (red, dashed line).  
The parameter Q has been marginalized.  For comparison, 
complete likelihood contours are shown in black (solid line) and
$\chi^2$ contours in green (dot--dashed line).  Note the improvement
due to the use of two parameters ($\dTnorm,m$) for the description 
of the MAX and Saskatoon bins.}
\end{figure}


\section{Conclusion}

     Among the various reasons for believing that
CMB temperature fluctuations are the cosmologist's
most useful tool for determining the fundamental 
Big Bang model parameters are their simple, {\em linear} physics 
(at least within the framework of standard passive 
perturbations) and straightforward interpretation.
By the latter we mean that there exists a very clear
connection between measured and theoretical
quantities -- an easily constructible likelihood function,
Eq. (\ref{eq:like1}).  Unfortunately, the computational complexity
of inverting the large covariance matrix $\mC$ and finding
its determinant many times
makes direct application of the likelihood approach
practically impossible.  It is really only possible to perform 
full likelihood treatments in pixel space for rather small 
data sets.  One way of improving the situation is to
work in the power plane of Figure 1, where sky 
temperatures are reduced to a much smaller number
of power estimates.  Gaussian fluctuations are
completely described by their multipole power spectrum,
so the data compression is lossless.  Real experiments
are however incapable, due to limited sky coverage and the
presence of noise, of recovering the individual multipoles.
This fact significantly complicates all CMB analysis
efforts.  The estimation of band--powers, their best 
definition and their use in statistical analyzes 
all become important, related and nontrivial issues.  
More specifically, one should address the question
of the adequacy of any method based on band--powers
to reproduce a full likelihood analysis.

     In this {\em paper} we have examined in detail some
of the current techniques applied to power estimates
to constrain cosmological parameters.  
To test the various methods, we have performed
a full likelihood analysis on the COBE, MAX
and Saskatoon experiments; this is a {\em prerequisite}
for making any solid statement concerning
the fidelity of an approximate method based on
power estimates.  We outlined  our likelihood 
analysis in Section 2.  Although necessarily 
restricted to a limited set of models (open scenarios
with zero cosmological constant), we nevertheless note
that interesting constraints are obtained with
this small data subset. 

  We found, perhaps not 
too surprisingly, that $\chi^2$ methods do not completely
reproduce the likelihood contours and are
susceptible to bias, all essentially due to 
their incorrect application of Gaussian 
distributions to power estimates.  Other methods
adopting more appropriate distributions 
fare better (Paper 1; Bond et al. 2000; Wandelt et
al. 2000).  More fundamentally, we found
that in certain situations flat band--powers 
do not always retain all the useful information 
of an experimental result.  This is
the case, for example, for some 
of the MAX and Saskatoon power bins. 
In these cases we found that the
data are more appropriately parameterized by
two spectral parameters, a local in--band 
power ($\dTnorm$) and a local in--band slope ($m$).
We were able to recover the lost
information by using these two parameters
rather than a single flat--band power estimate.

     In the present work, we have focused
on first generation data, for which
the observational strategy often suggests the
form of the adopted power band (e.g., via
the differencing scheme used). 
The issue concerning the fidelity of power estimates
is nevertheless of general importance.  
The number of power bands, or parameters of any kind,
used to summarize an observational campaign 
must correspond to the quantity of pertinent
information.  This may be addressed in practical
terms by testing to see how many independent parameters
(power bands or spectral parameters)
are significantly constrained by the pixel data.

     These issues, the faithfulness of both
the reduction to power estimates and the 
application of approximate power--based analysis
methods, will become more important as the
signal--to-noise of the observations increases,
and as one tries to extract ever more precise 
constraints from the data.  It will thus be of
all the more interest to find fast analysis techniques able
to reproduce as faithfully as possible complete
(but impossible to realize) likelihood results.



\end{document}